\documentclass{aastex}
\usepackage{spr-astr-addons}
\usepackage{url}
\pdfoutput=1

\newcommand{\emaila}{jdbruijn@rssd.esa.int}

\begin{document}

\title{Science performance of Gaia, ESA's space-astrometry mission}
\slugcomment{}
\shorttitle{Gaia status}
\shortauthors{De Bruijne}

\author{J.H.J. de Bruijne\altaffilmark{1}}
\altaffiltext{1}{European Space Agency (ESA), European Space Research and Technology Centre (ESTEC), Research and Scientific Support Department (SRE-SA), P.O. Box 299, 2200 AG, Noordwijk, the Netherlands, \emaila}

\begin{abstract}
Gaia is the next astrometry mission of the European Space Agency (ESA), following up on the success of the Hipparcos mission. With a focal plane containing $106$ CCD detectors, Gaia will survey the entire sky and repeatedly observe the brightest $1,000$ million objects, down to $20^{\rm th}$ magnitude, during its 5-year lifetime. Gaia's science data comprises absolute astrometry, broad-band photometry, and low-resolution spectro-photometry. Spectroscopic data with a resolving power of $11,500$ will be obtained for the brightest $150$ million sources, down to $17^{\rm th}$ magnitude. The thermo-mechanical stability of the spacecraft, combined with the selection of the L2 Lissajous point of the Sun-Earth/Moon system for operations, allows stellar parallaxes to be measured with standard errors less than $10$~micro-arcsecond ($\mu$as) for stars brighter than $12^{\rm th}$ magnitude, $25~\mu$as for stars at $15^{\rm th}$ magnitude, and $300~\mu$as at magnitude 20. Photometric standard errors are in the milli-magnitude regime. The spectroscopic data allows the measurement of radial velocities with errors of $15$~km~s$^{-1}$ at magnitude $17$. Gaia's primary science goal is to unravel the kinematical, dynamical, and chemical structure and evolution of the Milky Way. In addition, Gaia's data will touch many other areas of science, e.g., stellar physics, solar-system bodies, fundamental physics, and exo-planets. The Gaia spacecraft is currently in the qualification and production phase. With a launch in 2013, the final catalogue is expected in 2021. The science community in Europe, organised in the Data Processing and Analysis Consortium (DPAC), is responsible for the processing of the data.
\end{abstract}


\section{Science objectives}

The primary objective of the Gaia mission is to survey $1,000$ million stars in our Galaxy and beyond. Data from this census will foremost allow to address fundamental questions about the formation, structure, and evolution of our Galaxy, but will also provide unique insight into many other areas of astronomy.

\subsection{Galactic structure}

\begin{figure}[ht!]
\epsscale{1.0}
\plotone{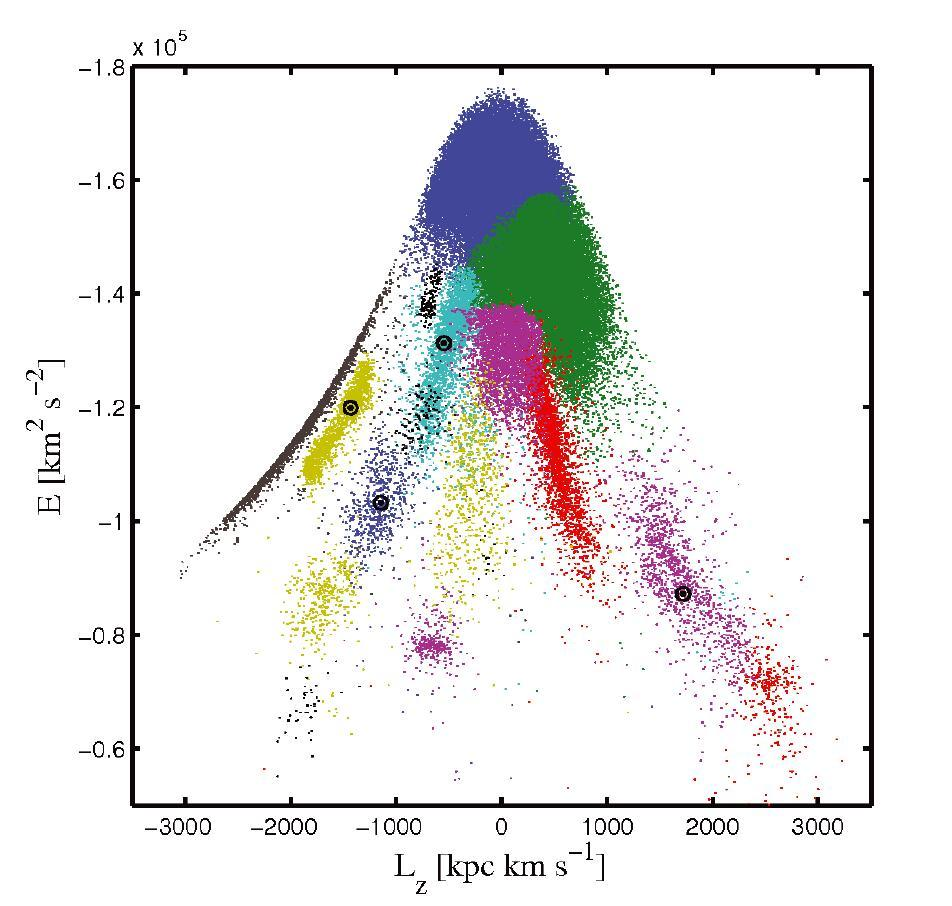}
\caption{\cite{2010MNRAS.408..935G} have simulated the formation of the Galactic stellar halo via the accretion of satellite galaxies onto a time-dependent, semi-cosmological, Galactic potential with the goal of characterising the sub-structure left by these accretion events in a manner to what will be possible with Gaia data. A synthetic solar-neighbourhood catalogue of stars was created by convolving the six-dimensional phase-space coordinates of stellar particles from disrupted satellites with the Gaia measurement errors. A realistic background contamination due to the Galactic discs and bulge was included. The resulting phase-space was found to be full of sub-structure, as exemplified by the clustered structure of coloured dots, each representative of a separate merger event, in the energy-angular momentum ($E$--$L_z$) space. With a newly developed clustering algorithm, \citeauthor{2010MNRAS.408..935G} were able to successfully isolate 50\% of the different satellites contributing to the solar neighbourhood; Fourier analysis of the space of orbital frequencies allowed them to deduce the times since accretion for 30\% of the recovered satellites (black open circles).}
\label{deBruijne_Figure_Halo}
\end{figure}

Gaia will perform a unique all-sky survey to map the three-dimensional position and velocity of all objects down to $20^{\rm th}$ magnitude. The Gaia data will encompass at least $1,000$ million stars that together cover a significant fraction of the Galaxy's volume: the accuracy and sensitivity of Gaia allows stars to be detected, and their position and velocity to be measured, from the solar neighbourhood, all the way through the disc of the Milky Way, to the bulge at the Galactic centre, and even further out into the halo. Accurate knowledge of stellar velocities and three-dimensional positions gives insight into the structure and dynamics of our Galaxy, including the build up of its different stellar components through past accretion and merger events with smaller satellite galaxies (Figure~\ref{deBruijne_Figure_Halo}).
 
The distributions of stars in the Galaxy over position and velocity are linked through gravitational forces and the star-formation rate and history as a function of position and time. The star-formation history can be derived from the observed population of stars by determining their distribution over stellar type (i.e., colour and luminosity). Since both luminosity and colour change in the course of a star's lifetime as it passes through different evolutionary stages, the position of a star in the Hertzsprung--Russell (colour-luminosity) diagram reveals its age. The observed distribution of the Galaxy's population of stars in the Hertzsprung--Russell diagram can, in principle, be compared with that of models containing collections of stars of different ages and colours. In practice, however, this method is limited in accuracy due to degeneracies in the effects of age and chemical composition on stellar colours and luminosities. Gaia's astrometric (distance), photometric (luminosity and extinction), and spectroscopic data (metallicity and abundances), combined with specifically developed tools, will resolve this ambiguity. Direct-inversion tools will make the evolutionary history of the Galaxy "directly" accessible.

Open questions on galactic structure that the Gaia data will be able to help answer are, for instance:
\begin{itemize}
\item Do galaxies form from accumulation of smaller systems, which have already initiated star formation? 
\item Does star formation begin in a gravitational potential well in which much of the gas is already accumulated? 
\item Does the bulge pre-date, post-date, or is it contemporaneous with the halo and inner disc? 
\item Is the thick disc a mix of the early disc and a later major merger? 
\item Is there a radial age gradient in the older stars? 
\item Is the history of star formation relatively smooth or highly episodic?
\end{itemize}

\subsection{Examples of miscellaneous science topics}

Some examples of miscellaneous science topics which will be addressed by Gaia are:
\begin{itemize}
\item For stars within $200$~pc from the Sun, Gaia will detect every Jupiter-size planet with an orbital period of $1.5$--$9$~years \citep{2011EAS....45..273S}. It will do this by revealing periodic shifts in the star's position, reflecting the gravitational pull of a planet in orbit around the star. Gaia is expected to astrometrically detect $2,000$ exo-planets, in addition to $5,000$ photometrically-detected "transiting" exo-planets.
\item Gaia will detect tens of thousands of brown dwarfs, both drifting through space in isolation and in orbit around other stars \citep{2002EAS.....2..199H}. This data is vital for investigating the physics of star formation since brown dwarfs represent stars that "just did not make it" to core hydrogen fusion.
\item Gaia will contribute to solar-system science because of its sensitivity to faint, moving objects \citep{2011EAS....45..225T}. Gaia will observe hundreds of thousands of minor planets and determine orbits with unprecedented accuracy. Although most of them will be main-belt asteroids, Gaia will also observe some near-Earth and Kuiper-Belt objects.
\item As star light passes by the Sun, or even a planet, large moon, or asteroid in the solar system, it is deflected by that object's gravitational field. Gaia will detect this shift and allow the most precise measurement of this general-relativistic effect ever, down to 2 parts in $10^6$ \citep{2010IAUS..261..306M}.
\item As shown in this {\it Astrophysics and Space Science} issue, Gaia will revolutionise the cosmic distance scale. Gaia will allow a parallax-based calibration of primary distance indicators, such as Cepheids and RR Lyrae stars. This, in turn, will allow a re-calibration of secondary indicators, such as type-Ia supernovae and globular-cluster luminosity functions, and hence a re-assessment of the entire distance~ladder.
\end{itemize}

\section{Mission and observational strategy}

Gaia will be launched from the European Space port in French Guiana by a Soyuz-STB/Fregat launch vehicle. Initially, the Fregat-Gaia composite will be placed in a low-altitude parking orbit, after which a Fregat boost will inject the spacecraft into its transfer trajectory towards the second Lagrange (L2) point of the Sun-Earth/Moon system. This point is located $1.5$ million km from Earth, in the anti-Sun direction, and co-rotates with the Earth in its one-year orbit around the Sun. After arriving at L2, one month after launch, the spacecraft will be inserted into its operational, Lissajous-type orbit around L2, with an orbital period of $180$~days and a size of $340,000$~km $\times$ $90,000$~km. Following commissioning and initial calibration, the spacecraft will be ready to enter the 5-year long nominal operational phase, which may be extended by one year. To ensure that Gaia stays close to L2, monthly orbit-maintenance manoeuvres are needed.

\begin{figure}[!t]
\epsscale{1.0}
\plotone{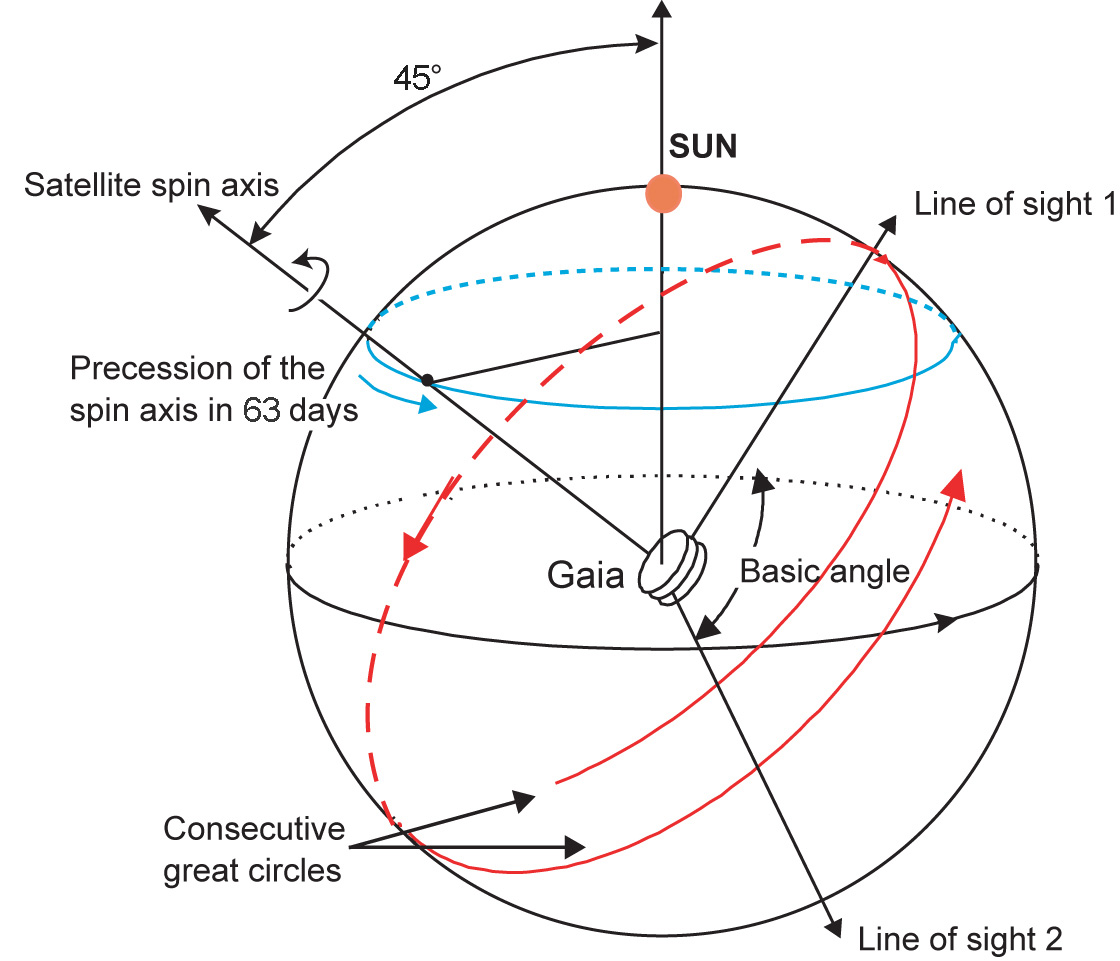}
\caption{Gaia's observing principle. L2 offers a stable thermal environment (Gaia's orbit around L2 is not impacted by Earth-induced eclipses of the Sun), a benign radiation environment, and a high observing efficiency since the Sun, Earth, and Moon are always outside the instrument fields of view. Uninterrupted sky mapping takes place during the 5-year operational mission phase. Image courtesy of ESA.}
\label{deBruijne_Figure_Principle}
\end{figure}

\begin{figure*}[!t]
\epsscale{1.5}
\plotone{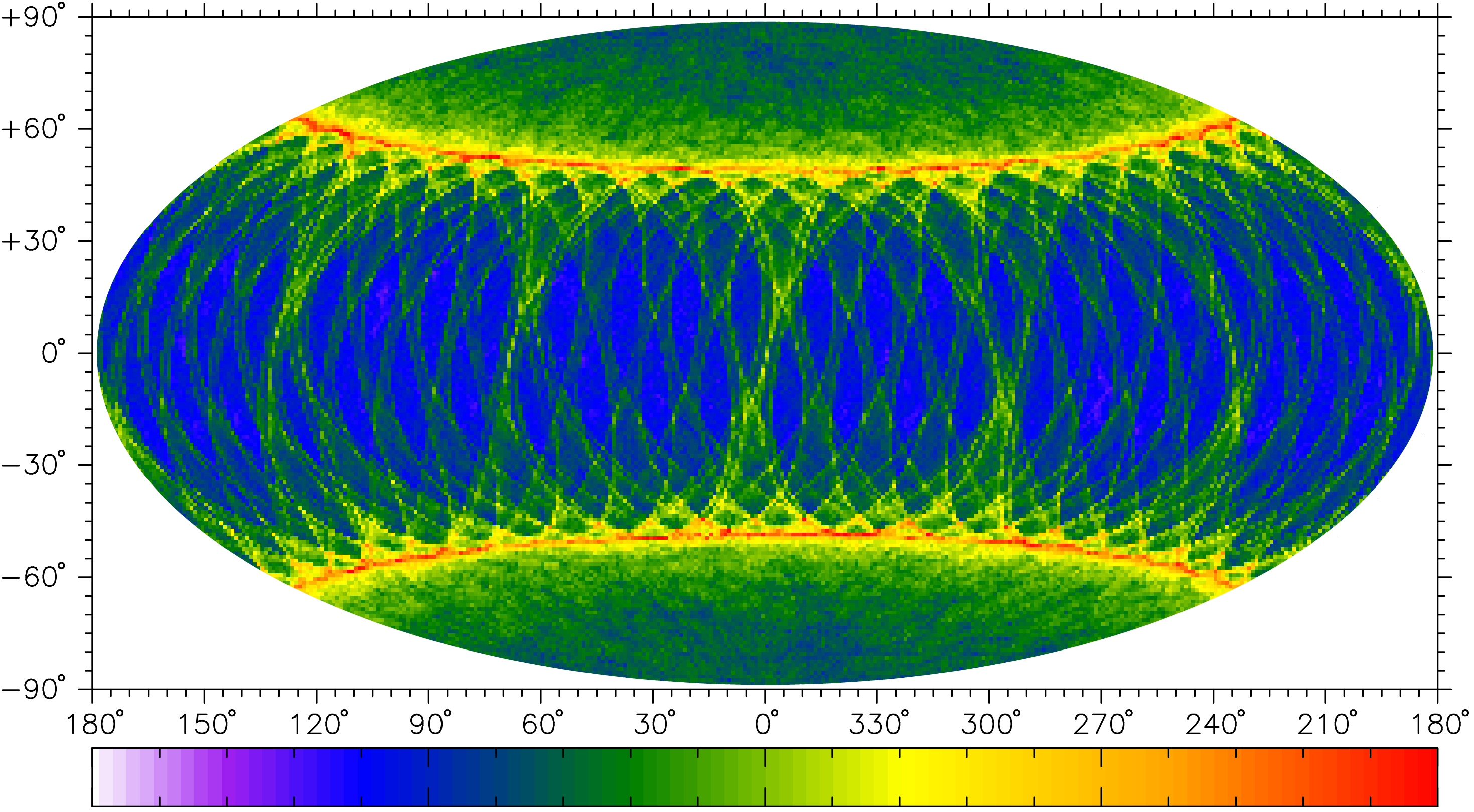}
\caption{End-of-mission sky coverage of the astrometric and photometric instruments, in ecliptic coordinates, in an equal-area Hammer--Aitoff projection. The colour wedge linearly spans the range from 0 (white) to 200 (red) focal-plane transits. The sky-average number of transits is 70, after accounting for unplanned spacecraft outages, data loss, etc. For the spectroscopic instrument, the map should be scaled by a factor 4/7. Image courtesy of ESA.}
\label{deBruijne_Figure_Sky_coverage}
\end{figure*}

Gaia builds on the global-astrometry concept successfully demonstrated by ESA's predecessor mission, Hipparcos \citep{1997A&A...323L..49P, 1997yCat.1239....0E, 2008yCat.1311....0V, 2010SSRv..151..209V}. This measurement principle (Figure~{\ref{deBruijne_Figure_Principle}) relies on the repeated observation of star positions in two fields of view. For this purpose, the spacecraft is rotating at a constant rate of $1^{\circ}$ per minute around an axis perpendicular to the two fields of view. With a basic angle of $106.5^{\circ}$ separating the fields of view, objects transit the second field of view $106.5$ minutes after crossing the first one. The spacecraft rotation axis makes an angle of $45^{\circ}$ with the Sun direction, representing the optimal choice between astrometric-performance requirements -- which favour a large angle -- and implementation constraints, such as payload shading and solar-array efficiency, which call for a small angle. The scan axis further describes a precession motion around the Sun-to-Earth direction, with a period of 63 days. The sky-average number of focal-plane transits over the lifetime is 70 (Figure~\ref{deBruijne_Figure_Sky_coverage}).

\begin{figure}[t!]
\epsscale{1.00}
\plotone{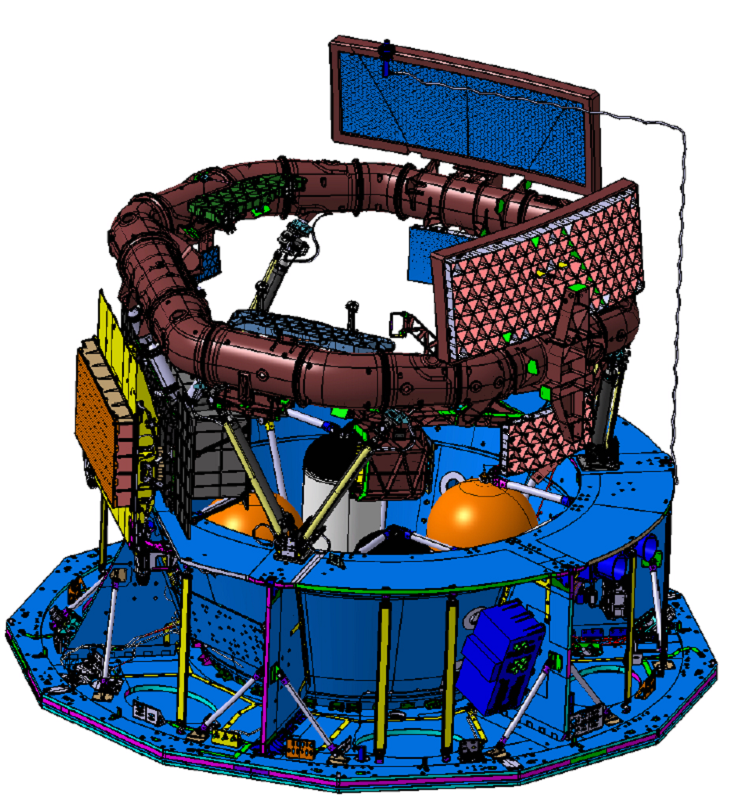}
\caption{Drawing of the Gaia spacecraft. The payload module with the optical bench is depicted in brown. It rests -- by means of three bipods -- on the service module, which is the blue structure in the lower half of the image. The focal-plane assembly is the box hanging from the optical bench on the left. The two M1/M1$^\prime$ primary mirrors of the two telescopes are mounted on top of the optical bench. The micro- and chemical-propulsion tanks are visible inside the service module. Not depicted are the thermal tent, which covers the service and payload modules, and the deployable sunshield assembly. The vertical line at the right edge is the connection to the low-gain antenna placed on top of the thermal tent (visible in front of the primary mirror in the back). Image courtesy of EADS Astrium SAS, France.}
\label{deBruijne_Figure_Spacecraft}
\end{figure}

\section{Spacecraft}

The Gaia spacecraft (built under responsibility of EADS Astrium SAS, France; Figure~\ref{deBruijne_Figure_Spacecraft}) is comprised of a payload module (built under responsibility of EADS Astrium SAS, France), a mechanical service module (built under responsibility of EADS Astrium GmbH, Germany), and an electrical service module (built under responsibility of EADS Astrium Ltd, United Kingdom). The payload module is built around an optical bench which provides structural support for the telescopes and instruments. It further contains all necessary electronics for managing the instrument operation and processing and storing the raw data. The mechanical service module comprises all mechanical, structural, and thermal elements supporting the instrument and the spacecraft electronics. It also includes the chemical and micro-propulsion systems, deployable-sunshield assembly, thermal tent, solar-array panels, and electrical harness. The electrical service module offers support functions to the payload and spacecraft for pointing, electrical power control and distribution, central data management, and communications.

\subsection{Telescope and optical bench}

\begin{figure*}[t!]
\epsscale{2.29}
\plottwo{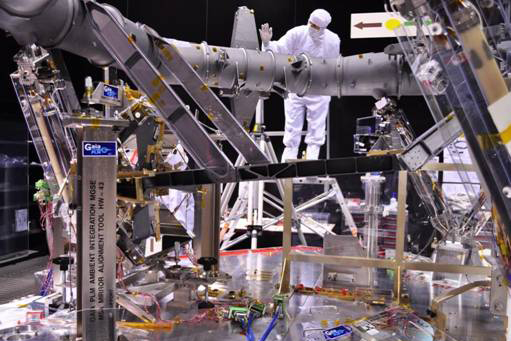}{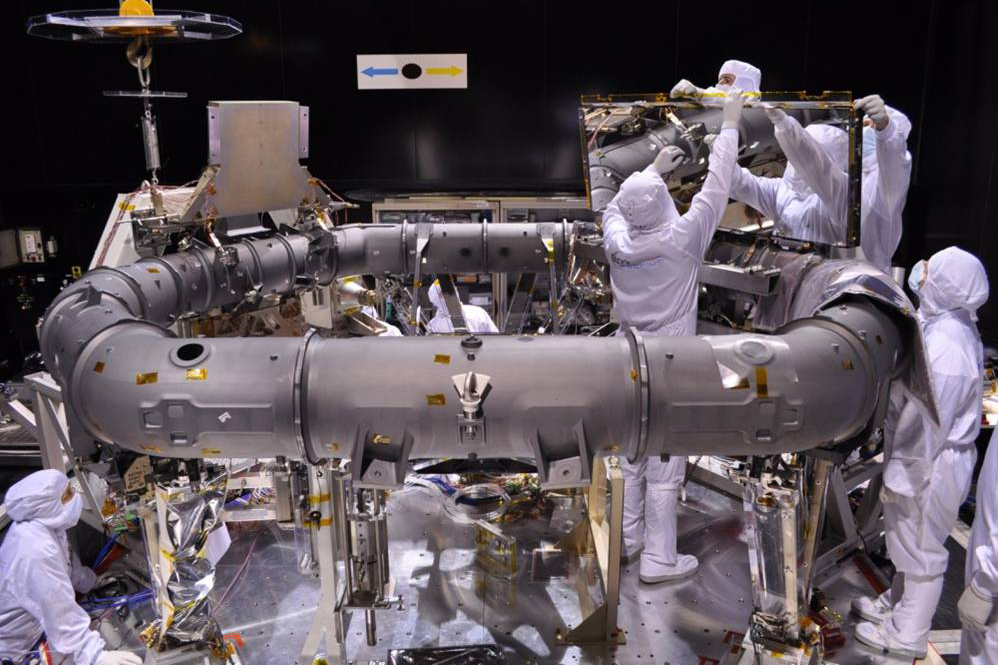}
\caption{The payload module under integration in the cleanroom. Left: the M5 and M6 $45^{\circ}$ flat folding mirrors are visible on the left. The M4/M4$^\prime$ beam-combiner mirrors are visible to the right of the legs of the person. Right: integration of the M1 primary mirror. Images courtesy of EADS Astrium SAS, France.}
\label{deBruijne_Figure_Torus}
\end{figure*}

Gaia's payload carries two identical telescopes, pointing in two directions separated by a $106.5^{\circ}$ basic angle and merged into a common path at the exit pupil. The optical path of both telescopes is composed of six reflectors (M1--M6), the first three (M1--M3 and M1$^\prime$--M3$^\prime$) forming two three-mirror anastigmats, while the last two are a common-path beam folder (M5--M6). Beam combination is achieved in image space with a beam combiner, at M4/M4$^\prime$ level. High reflectivity on all mirrors is achieved by a protected silver coating. Both telescopes have an aperture of $1.45$~m $\times$ $0.50$~m and a focal length of $35$~m. The telescopes are mounted on a hexagonal optical bench (Figure~\ref{deBruijne_Figure_Torus}). The optical bench and all telescope mirrors are built of silicon-carbide (SiC). This is an ultra-stable material, providing low mass, high isotropy and thermo-elastic stability, and optimum dimensional stability. The use of SiC allows meeting the stringent stability requirements for the basic angle between the two telescopes with passive rather than active thermal control.

\subsection{Focal-plane assembly}

The Gaia focal-plane assembly \citep{2007SPIE.6690E...8L} is the largest ever developed for space application, with 106 CCDs (Figure~\ref{deBruijne_Figure_FPA}), a total of almost $1,000$ million pixels, and a physical dimension of $1.0$~m $\times$ $0.5$~m. The focal-plane assembly is common to both telescopes and serves five main functions:
\begin{itemize}
\item The wave-front sensor \citep{2009SPIE.7439E..29V} and basic-angle monitor \citep{2009SPIE.7439E..30M}, covering $2 + 2$ CCDs: a five-degrees-of-freedom mechanism is implemented behind the M2/M2$^\prime$ secondary mirrors of the two telescopes for re-aligning the telescopes in orbit to cancel errors due to mirror micro-settings and gravity release. These devices are activated following the output of two Shack--Hartmann-type wave-front sensors at different positions in the focal plane. The basic-angle monitor system (two CCDs in cold redundancy) consists of a Youngs-type interferometer continuously measuring fluctuations in the basic angle between the two telescopes with a resolution of $0.5$~$\mu$as per 5 minutes;
\item The sky mapper (SM), containing 14 CCDs (seven per telescope), which autonomously detects objects down to $20^{\rm th}$ magnitude entering the fields of view and communicates details of the star transits to the subsequent CCDs;
\item The main astrometric field (AF), covering 62 CCDs, devoted to ang\-ular-position measurements, providing the five astrometric parameters: star position (two angles), proper motion (two time derivatives of position), and parallax (distance) of all objects down to $20^{\rm th}$ magnitude. The first strip of seven detectors (AF1) also serves the purpose of object confirmation;
\item The blue and red photometers (BP and RP), providing low-resolution, spectro-photometric measurements for each object down to 20$^{\rm th}$ magnitude over the wavelength ranges $330$--$680$~nm and $640$--$1050$~nm, respectively. The data serves general astrophysics and enables the on-ground calibration of telescope-induced chromatic image shifts in the astrometry\footnote{Although the optical design is exclusively based on mirrors, diffraction effects with residual aberrations induce systematic chromatic shifts of the diffraction images, and thus of the measured star positions. These chromatic displacements, usually neglected in optical systems, are significant for Gaia and will be calibrated as part of the on-ground data analysis using the colour information provided by the photometry of each observed object.}. The photometers contain seven CCDs each;
\item The radial-velocity spectrometer (RVS), covering 12 CCDs in a 3 $\times$ 4 arrangement, collecting high-resolution spectra of all objects brighter than $17^{\rm th}$ magnitude, allowing derivation of radial velocities and stellar atmospheric parameters.
\end{itemize}

\begin{figure*}[t!]
\epsscale{2.29}
\plottwo{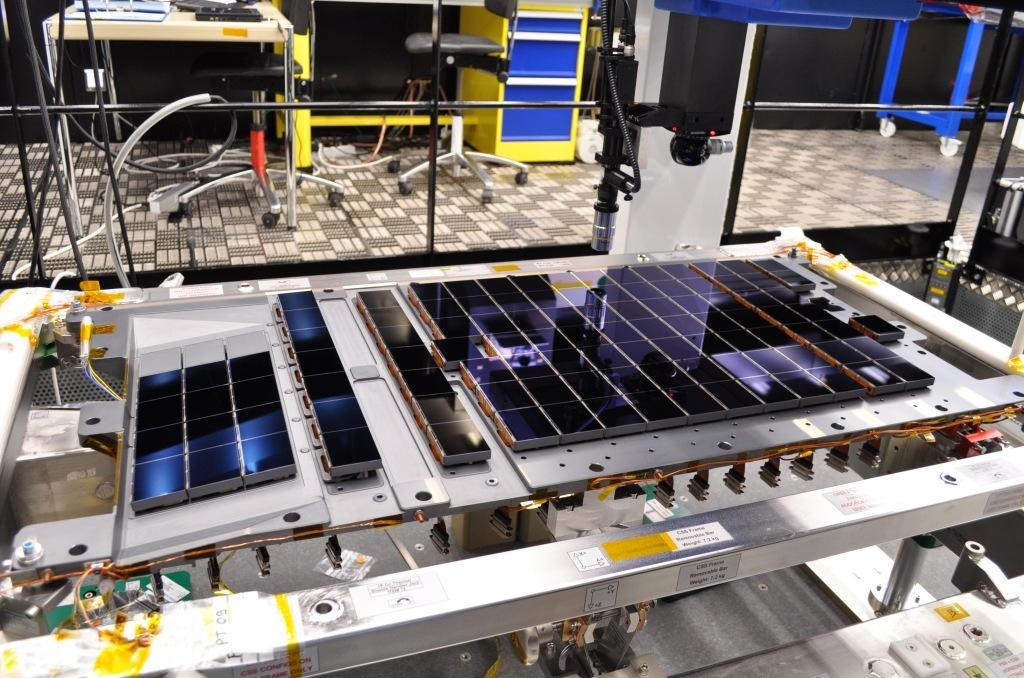}{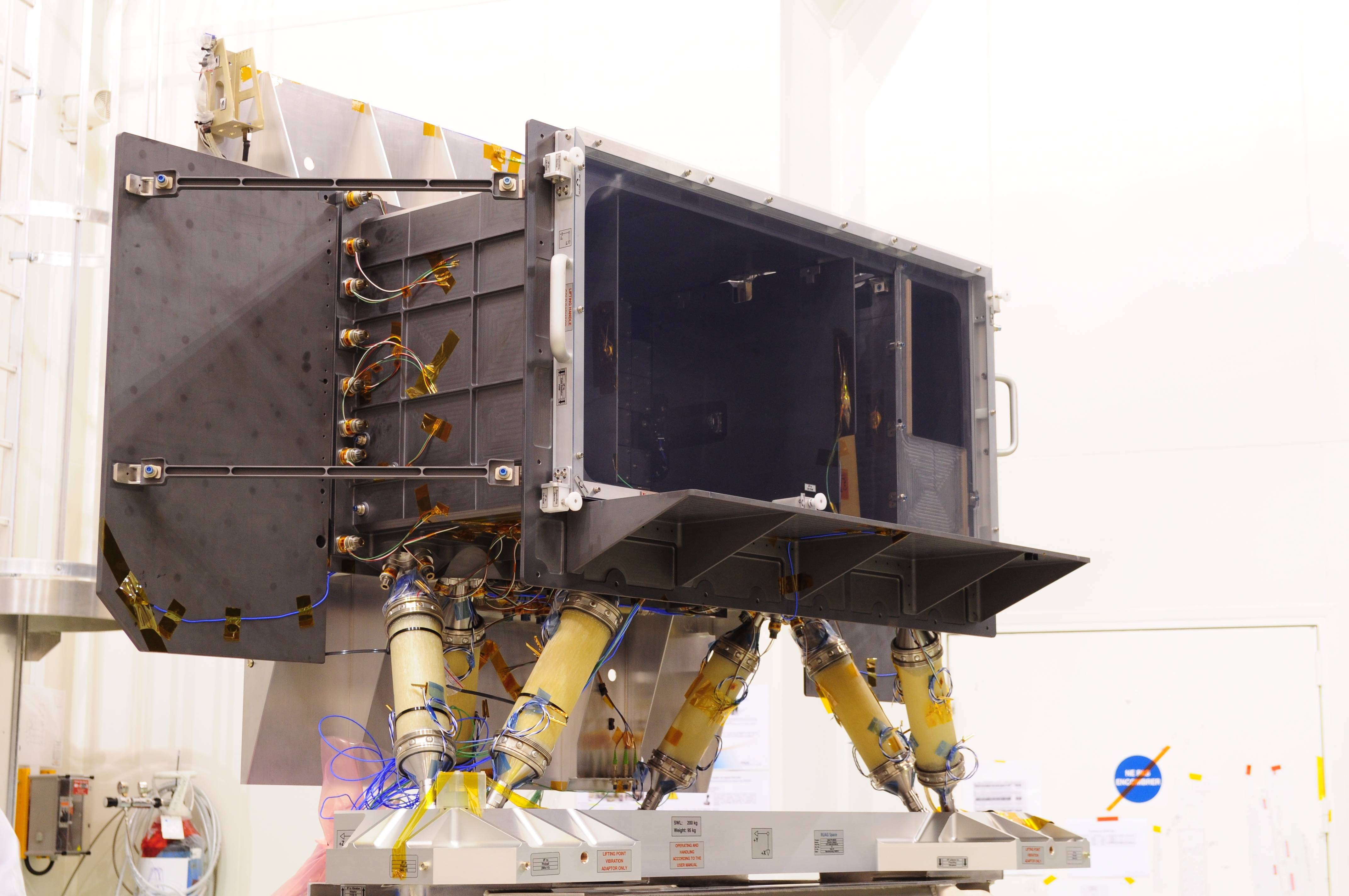}
\caption{Left: 106 CCD detectors make up Gaia's focal plane. The CCDs have been manufactured by e2v technologies of Chelmsford, United Kingdom. Each CCD measures $4.5$~cm $\times$ $5.9$~cm; the associated field-of-view on the sky is $4.4^\prime$ $\times$ $5.8^\prime$. The focal plane is passively cooled to $170$~K for negligible dark current and for minimising the radiation-induced charge-transfer inefficiency \citep{2010SPIE.7742E..28S,2010SPIE.7742E..30S,2011MNRAS.414.2215P,2011PhDT.........9P}. Right: the focal-plane-assembly structural model in the cleanroom, prior to mechanical and structural testing. The box-shaped cold radiator provides the radiative surface with the internal payload cavity, as well as shielding against radiation and mounting support for the photometer prisms. The focal plane itself is inside the radiator. The six yellow "tubes" are the support struts to mount the assembly on the optical bench. Images courtesy of EADS Astrium SAS, France.}
\label{deBruijne_Figure_FPA}
\end{figure*}

All CCDs, except those in the sky mapper, are operated in windowing mode: only those parts of the CCD data stream which contain objects of interest are read out; remaining pixel data is flushed at high speed. The use of windowing mode reduces the readout noise to a handful of electrons while still allowing reading up to 20 objects simultaneously.

Every object crossing the focal plane is first detected either by SM1 or SM2. These CCDs record, respectively, the objects only from telescope 1 or from telescope 2. This is achieved by a physical mask that is placed in each telescope intermediate image, at M4/M4$^\prime$ beam-combiner level. Next, the window is allocated to the object, which is propagated through the following CCDs of the CCD row as the imaged object crosses the focal plane; the actual propagation uses input from the spacecraft's attitude control system, which provides the predicted position of each object in the focal plane versus time. After detection in SM, each object is confirmed by the CCD detectors in the first strip of the astrometric field (AF1); this step eliminates false detections such as cosmic rays. The object then progressively crosses the eight next CCD strips in AF, followed by the BP, RP, and RVS detectors (the latter are present only for four of the seven CCD rows).

\subsection{CCD detectors}

All CCDs in the focal plane are the same model, the e2v technologies CCD91-72 \citep{2008SPIE.7106E..40W, 2009SPIE.7439E...9K}. The detectors are back-illuminated devices with an image area of 4500 lines $\times$ 1966 columns of 10 $\mu$m $\times$ 30 $\mu$m pixel size, a compromise to achieve high resolution in the along-scan direction as well as sufficient pixel-full-well capacity. All CCDs are operated in time-delay-integration (TDI) mode with a TDI period of $982.8$~$\mu$s, synchronised with the spacecraft scanning motion. The integration time per CCD is $4.42$~seconds, corresponding to the 4500 TDI lines along-scan. At distinct positions over the 4500 lines, a set of 12 special, electronic gates (TDI gates) is implemented, which can be used to temporarily or permanently block charge transfer over these lines and hence effectively reduce the TDI integration time. While the full 4500-lines integration is used for faint objects, the activation of the special gates for bright objects permits avoiding star-image saturation at CCD-pixel level and hence extends the detection limit towards brighter stars.

\begin{figure*}[t!]
\epsscale{2.0}
\plotone{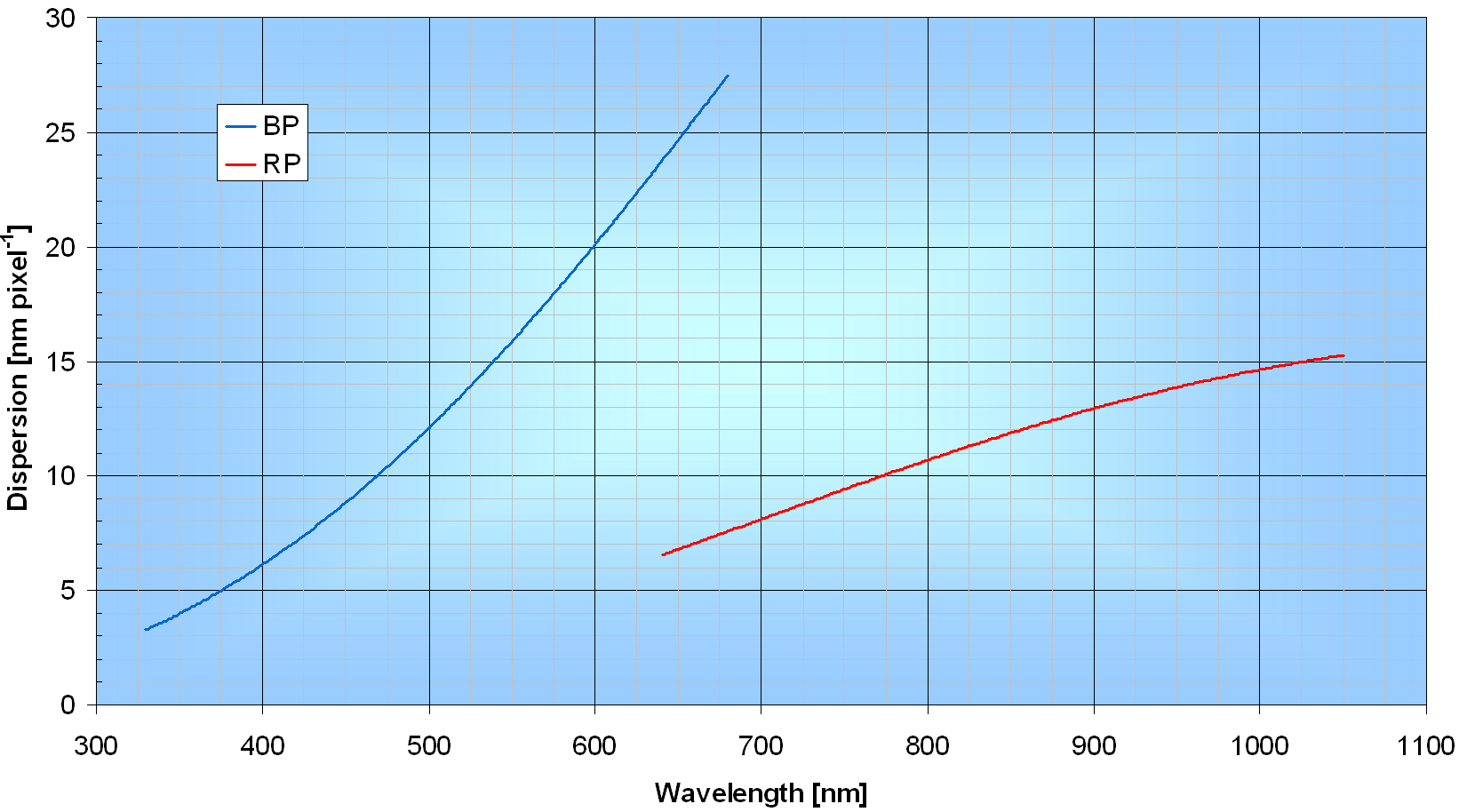}
\caption{The spectral dispersion of the photometric instrument is a function of wavelength and varies in BP from 3 to 27~nm pixel$^{-1}$ covering the wavelength range $330$--$680$~nm. In RP, the wavelength range is $640$--$1050$~nm with a spectral dispersion of 7 to 15~nm pixel$^{-1}$. The 76\%-energy extent of the line-spread function along the dispersion direction varies along the BP spectrum from 1.3 pixels at 330 nm to 1.9 pixels at 680 nm and along the RP spectrum from 3.5 pixels at 640 nm to 4.1 pixels at 1050 nm. Image courtesy of ESA.}
\label{deBruijne_Figure_XP_dispersion}
\end{figure*}

Gaia CCDs are fabricated in three variants -- AF-, BP-, and RP-type -- to optimise quantum efficiency corresponding to the different wavelength ranges of the scientific functions. The AF-type variant is built on standard silicon with broad-band anti-reflection coating. It is the most abundant type in the focal plane, used for all but the photometric and spectroscopic functions. The BP-type only differs from the AF-type through the blue-enhanced backside treatment and anti-reflection coating, and it is exclusively used in BP. The RP-type is built on high-resistivity silicon with red-optimised anti-reflection coating to improve near-infrared response. It is used in RP as well as in RVS.

\subsection{Astrometric instrument}

The main objective of the astrometric instrument is to obtain accurate measurements of the positions and velocities of all objects that cross the fields of view of the two telescopes. In essence, Gaia continuously measures the instantaneous relative separations of the thousands of stars simultaneously present in the two fields. The spacecraft operates in a continuously scanning motion, such that a constant stream of relative angular measurements is built up as the fields of view sweep across the sky (Figure~\ref{deBruijne_Figure_Sky_coverage}). The full set of relative measurements permit, after ground processing, a complete determination of each star's five astrometric parameters: two specifying the angular position, two specifying the proper motion, and one -- parallax -- specifying the star's distance. The astrometric data also permits determination of additional parameters, for example those relevant to orbital binaries, exo-planets, and solar-system objects. High angular resolution -- and hence high positional precision -- in the scanning direction is provided by the large primary mirror of each telescope. The wide-angle measurements provide high rigidity of the resulting reference system. The accuracy of the astrometric measurements -- in particular the zero point of the parallaxes -- critically relies on (knowledge and calibration of) the stability of the basic angle between the two telescopes.

The astrometric instrument comprises an area of 62 CCDs in the focal plane where the two fields of view are combined onto the astrometric field. Stars entering the field of view first pass across the strip of sky-mapper CCDs, where each object is autonomously detected by the on-board image-detection software. Information on an object's position and brightness is processed on board in real-time by the video-processing unit in order to define the windowed region around the object to be read out by the following CCDs in the focal plane. The astrometric data are binned on-chip in the across-scan direction over 12 pixels. For bright stars, single-pixel-resolution windows are used, in combination with TDI gates to shorten CCD integration times and hence avoid pixel-level saturation. The astrometric instrument can handle densities of $1,000,000$~objects deg$^{-2}$.

\begin{figure*}[t!]
\epsscale{2.0}
\plotone{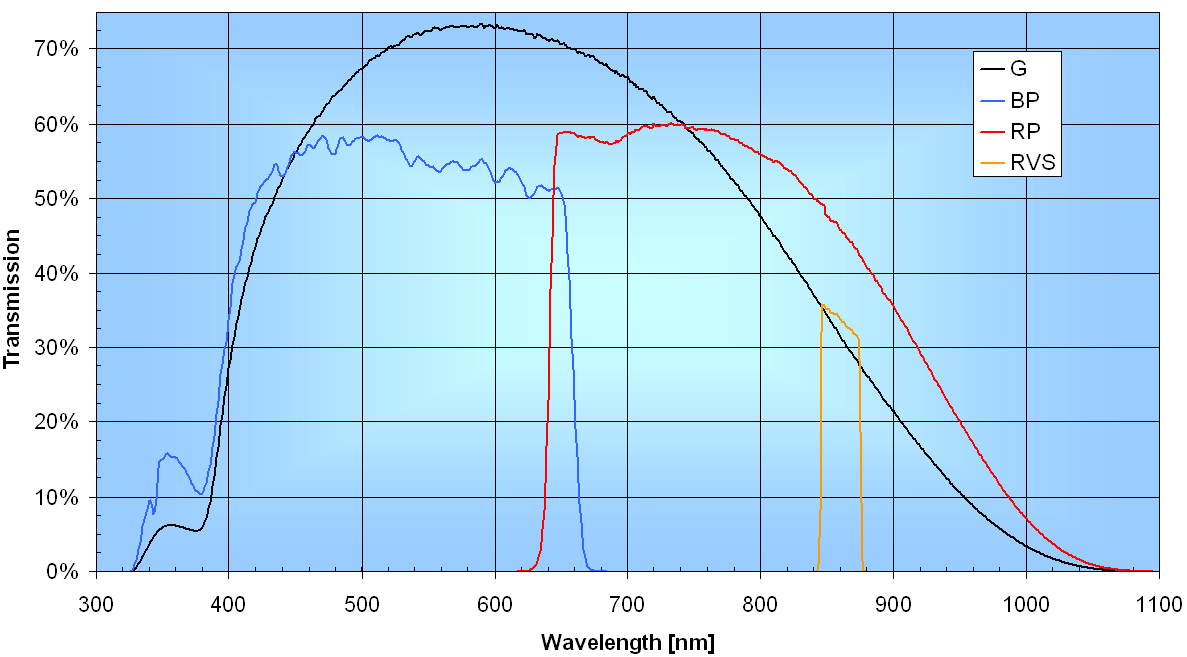}
\caption{Effective bandpasses of the astrometric (SM/AF), photometric (BP/RP), and spectroscopic (RVS) instruments, including telescope transmission, CCD quantum efficiency, and throughput of dispersing optics. \cite{2010A&A...523A..48J} provide colour transformations between Gaia, Johnson--Cousins, and Sloan passbands. Image courtesy of ESA.}
\label{deBruijne_Figure_Transmission}
\end{figure*}

\begin{figure}[ht!]
\epsscale{1.0}
\plotone{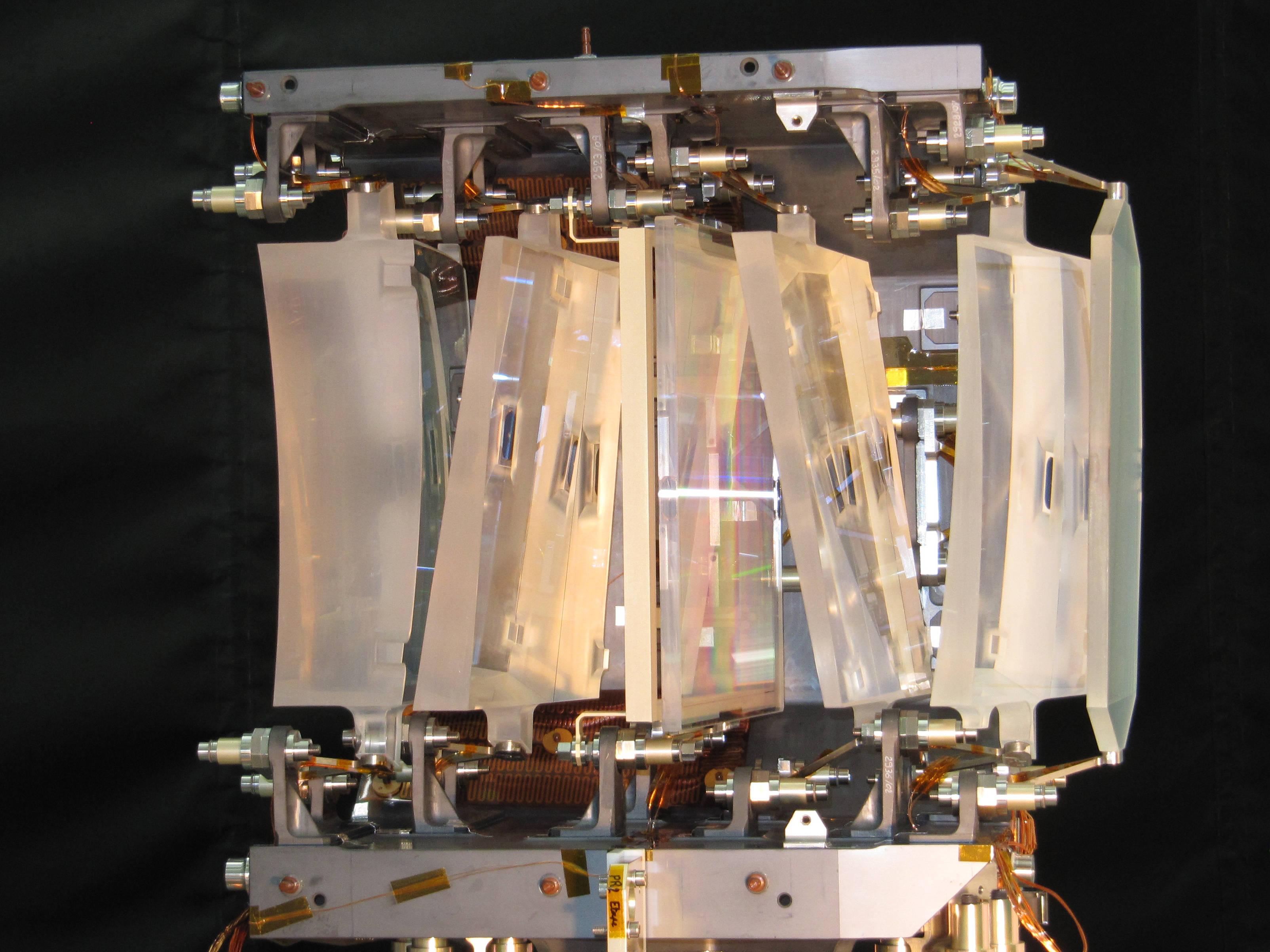}
\caption{The optical module of the radial-velocity spectrometer, containing a grating plate (middle), four fused-silica prismatic lenses, as well as a bandpass-filter plate (far right). Image courtesy of EADS Astrium SAS, France.}
\label{deBruijne_Figure_RVS_OMA}
\end{figure}

\subsection{Photometric instrument}

\begin{figure*}[t!]
\epsscale{2.0}
\plotone{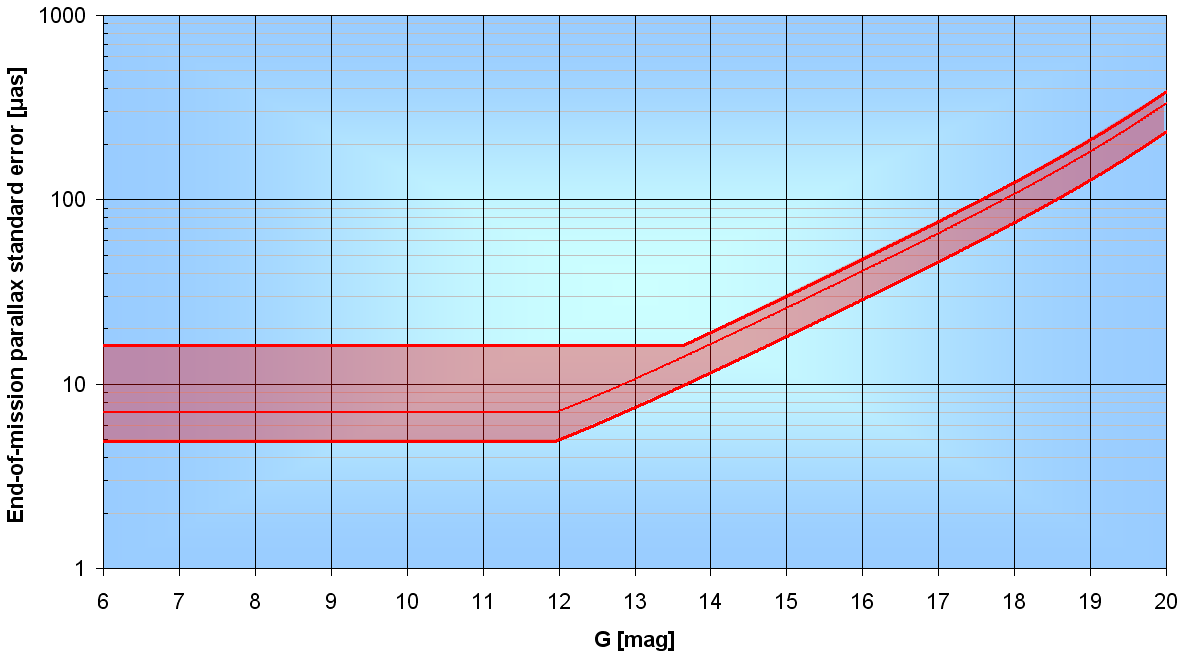}
\caption{Sky-average, end-of-mission parallax standard error, in $\mu$as, as function of Gaia's broad-band $G$ magnitude, for an unreddened G2V star with $V-I = 0.75$~mag and $V-G = 0.16$~mag. A functional parametrisation of this curve as function of magnitude and colour can be found on \protect\url{http://www.rssd.esa.int/index.php?project=GAIA&page=Science_Performance}. The standard-error calculation includes all known instrumental effects as well as a 20\% science margin. The bright-star errors -- which depend on the adopted TDI-gate scheme, which is not yet frozen and configurable in flight, as well as on magnitude -- have been set to a constant noise floor. The shaded area reflects the range resulting from varying the sky position, the $V-I$ colour / spectral type, and the bright-star (TDI-gate) observing strategy. Image courtesy of~ESA.}
\label{deBruijne_Figure_Astrometric_performance}
\end{figure*}

The photometric instrument measures the spectral energy distribution of detected objects, to allow derivation of astrophysical quantities such as luminosity, effective temperature, and chemical composition, as well as to enable astrometric calibration of telescope-aberration-induced chromatic shifts (see footnote~1). The photometric instrument (like the spectroscopic instrument) is merged with the astrometric function, using the same collecting apertures of the two telescopes. The photometry function is achieved by means of two low-dispersion, fused-silica prisms located in the common path of the two telescopes: one for short wavelengths (BP, the Blue Photometer, covering $330$--$680$~nm) and one for long wavelengths (RP, Red Photometer, covering $640$--$1050$~nm).

Both prisms are mounted on the CCD cold radiator, close to the focal plane, in order to reduce the shadow size on the focal plane. Both photometers have a CCD strip of seven CCDs each that covers the full astrometric field of view in the across-scan direction. The object-handling capability of the photometric instruments is limited to $750,000$~objects deg$^{-2}$.

The prisms disperse star images along the scan direction and spread them over 45 pixels. The spectral resolution is a function of wavelength as a result of the natural dispersion characteristics of fused silica (Figure~\ref{deBruijne_Figure_XP_dispersion}). BP and RP spectra are binned on-chip in the across-scan direction over 12 pixels. For bright stars, single-pixel-resolution windows are used, in combination with TDI gates to shorten CCD integration times and hence avoid pixel-level saturation.

\subsection{Spectroscopic instrument}

The primary objective of Gaia's radial-velo\-ci\-ty-spectro\-meter (RVS) instrument is the acquisition of radial velocities. These complement the proper-motion measurements provided by the astrometric instrument and hence provide three-dimensional stellar velocities. RVS collects spectra in the wavelength range\footnote{This range has been carefully selected to coincide with the energy-distribution peaks of G- and K-type giants, which are the most abundant RVS targets. For these late-type stars, the RVS wavelength range contains, besides numerous weak lines mainly due to Fe, Si, and Mg, three strong ionised-calcium lines. The absorption lines in this triplet allow radial velocities to be derived, even at modest signal-to-noise ratios. In early-type stars, RVS spectra contain weak lines such as Ca II, He I, He II, and N I, although they are dominated by hydrogen-Paschen lines.} $847$--$874$~nm (Figure~\ref{deBruijne_Figure_Transmission}) with a resolving power of $11,500$.

\begin{figure*}[t!]
\epsscale{2.0}
\plotone{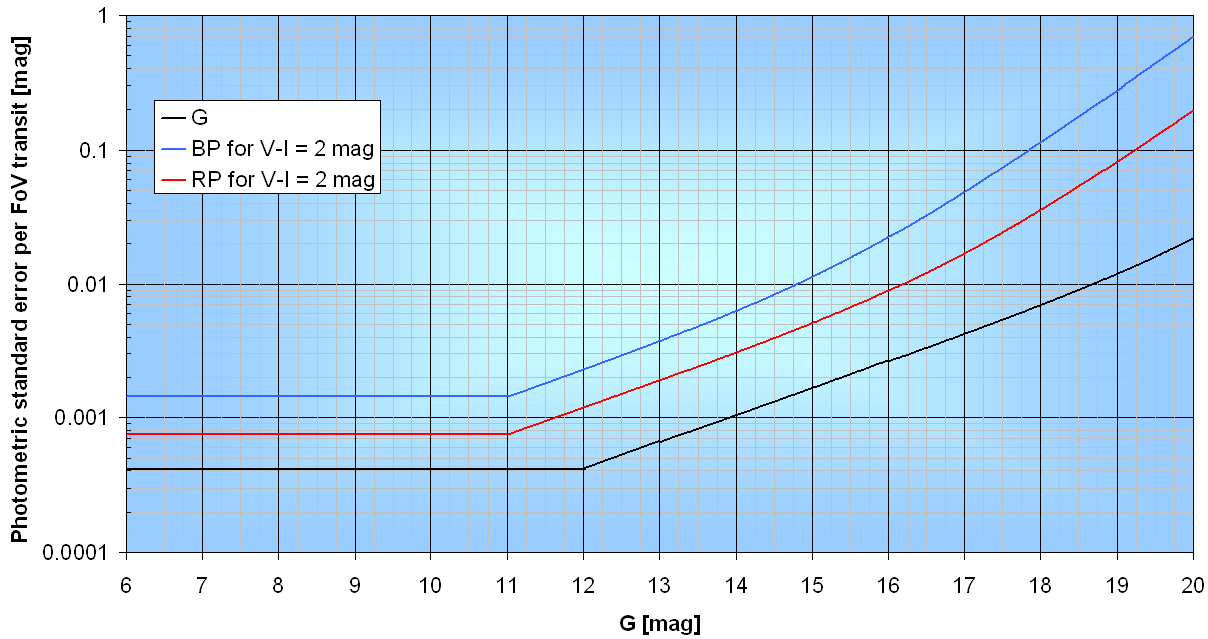}
\caption{Single-transit, photometric standard errors, in mag, as function of $G$ magnitude, for a star with $V-I = 2$~mag, for the integrated $G$-, BP-, and RP-band. A functional parametrisation of this curve as function of magnitude and colour can be found on \protect\url{http://www.rssd.esa.int/index.php?project=GAIA&page=Science_Performance}. The standard-error calculation includes all known instrumental effects as well as a 20\% science margin. The bright-star errors -- which depend on the adopted TDI-gate scheme, which is not yet frozen and configurable in flight, as well as on magnitude -- have been set to a constant noise floor. Assuming calibration errors are negligibly small, end-of-mission errors can be estimated by division of single-transit errors by the square root of the number of transits (70 in average -- Figure~\ref{deBruijne_Figure_Sky_coverage}). Image courtesy of~ESA.}
\label{deBruijne_Figure_Photometric_performance}
\end{figure*}

RVS is an integral-field spectrograph dispersing all light entering the field of view. Spectral dispersion -- oriented in the along-scan direction -- is achieved through an optical module (Figure~\ref{deBruijne_Figure_RVS_OMA}), located between telescope mirror M6 and the focal plane. This module contains a grating plate and four prismatic lenses to correct the main aberrations of the off-axis field of the telescope. A bandpass filter restricts the throughput of the RVS to the required wavelength range.

RVS is integrated with the astrometric and photometric functions and uses the common path of the two telescopes and focal plane. Objects are selected for RVS observation based on the RP spectral measurements made slightly earlier. The RVS part of the focal plane contains three CCD strips and four CCD rows, and each source will hence be observed during 40 focal-plane transits (120 CCD transits -- Figure~\ref{deBruijne_Figure_Sky_coverage}) throughout the mission. On the sky, the RVS CCD rows are aligned with the astrometric and photometric CCD rows; the resulting semi-simultaneity of the astrometric, photometric, and spectroscopic transit data is advantageous for stellar variability analyses, scientific alerts, spectroscopic binaries, etc. RVS spectra are binned on-chip in the across-scan direction over 10 pixels, except for bright stars, for which single-pixel-resolution windows are used. All single-CCD spectra are transmitted to ground without on-board processing. RVS can handle densities up to $36,000$~objects deg$^{-2}$ -- see \cite{2005MNRAS.359.1306W} for a discussion on crowding.

\section{Science performance}

\subsection{Astrometry}

\begin{figure*}[t!]
\epsscale{2.0}
\plotone{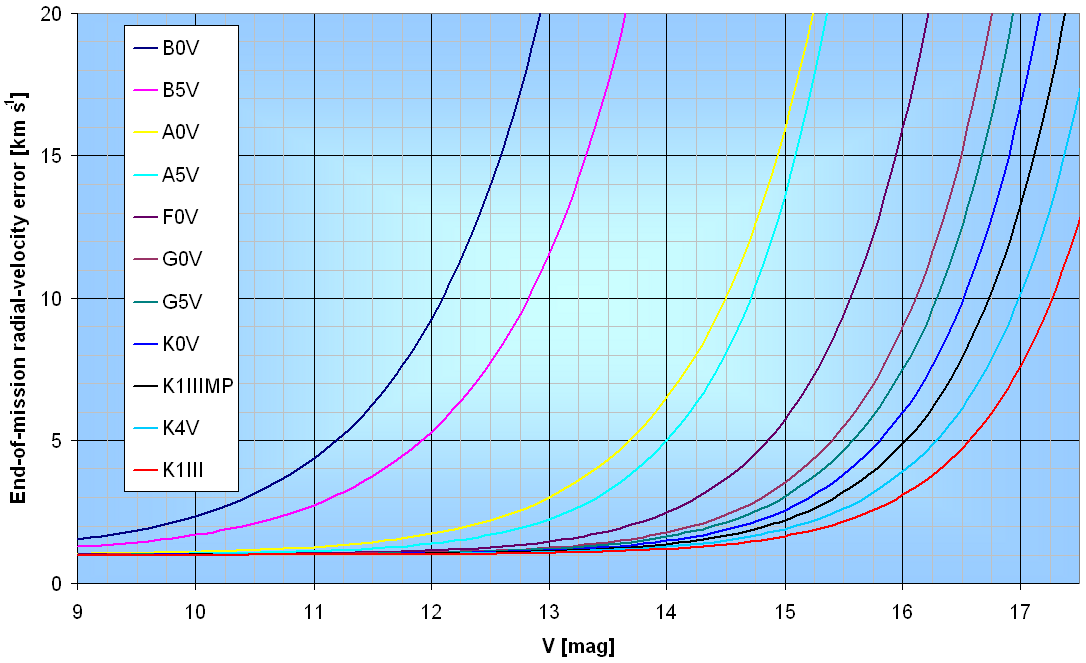}
\caption{Sky-average, end-of-mission radial-velocity standard errors, in km~s$^{-1}$, for a variety of stars as function of Johnson $V$ magnitude. Functional parametrisations of these curves as function of magnitude can be found on \protect\url{http://www.rssd.esa.int/index.php?project=GAIA&page=Science_Performance}. The error calculation includes all known instrumental effects as well as a 20\% science margin. Figure based on the DPAC-internal study report GAIA-C6-TN-OPM-PS-006 by Paola Sartoretti and collaborators. Image courtesy of ESA.}
\label{deBruijne_Figure_RVS_performance}
\end{figure*}

The on-ground processing of the astrometric data \citep{2011EAS....45..123L,2011arXiv1112.4139L} is a complex task, linking all relative measurements  and transforming the location (centroiding) measurements in pixel coordinates to angular-field coordinates through a geometrical calibration of the focal plane, and subsequently to coordinates on the sky through calibrations of the instrument attitude and basic angle. Further corrections to be performed on ground include those for systematic chromatic shifts (footnote~1) and general-relativistic effects (light bending due to the Sun, the major planets plus some of their moons, and the most massive asteroids).

The wavelength coverage of the astrometric instrument, defining the white-light, broad-band $G$ band, is $330$--$1050$~nm (Figure~\ref{deBruijne_Figure_Transmission}). The astrometric science performance is normally quantified by the end-of-mission parallax standard error. This error depends on $G$ magnitude as shown in Figure~\ref{deBruijne_Figure_Astrometric_performance} (the error also depends on position on the sky through the scanning law with which Gaia scans the sky -- Figure~\ref{deBruijne_Figure_Sky_coverage}; performance numbers reported here refer to sky-average values). Sky-averaged position errors are a factor 0.74 smaller than parallax errors; sky-averaged annual-proper-motion errors are a factor 0.53 smaller than parallax errors.

Due to their extreme brightness, Gaia will not be able to observe the 6,000 brightest stars in the sky, those with $G < 6$~mag. For stars fainter than $G = 6$~mag yet brighter than $G = 12$~mag, data {\it will} be acquired: shorter CCD integration times (through the use of TDI gates) will be used to avoid saturation. For these stars, the end-of-mission performance depends on the adopted TDI-gate scheme, which is not yet frozen and configurable in flight, as well as on $G$ magnitude.

\subsection{Photometry}

Whereas the $G$-band photometric data associated with the astrometric data are particularly useful for stellar variability studies \citep{2011EAS....45..161E}, the BP/RP spectra -- sometimes in combination with the astrometric and the spectroscopic data -- allow to retrieve astrophysical parameters of objects and to classify them \citep{2011EAS....45..381B}. Despite the non-applicability to Gaia's low-resolution, spectro-photometric data, the usual calculation methodology for photometric standard errors is based on an aperture-photometry paradigm. Typical resulting performances of the integrated $G$-, BP-, and RP-band are shown in Figure~\ref{deBruijne_Figure_Photometric_performance}.

The relevant quantities for the exploitation of Gaia's photometric data are not band fluxes but astrophysical parameters, such as interstellar extinctions and surface gravities. \cite{2010MNRAS.403...96B} shows that, for unreddened stars covering a wide range of metallicity, surface gravity, and effective temperature, BP/RP photometry allows to estimate $T_{\rm eff}$ to an accuracy of 0.3\% at $G = 15$~mag and 4\% at $G = 20$~mag. [Fe/H] and log($g$) can be estimated to accuracies of 0.1--0.4 dex for stars with $G \leq 18.5$~mag, depending on the magnitude and on what priors can be placed on the astrophysical parameters. If extinction varies {\it a priori} over a wide range, log($g$) and [Fe/H] can still be estimated to 0.3 and 0.5 dex, respectively, at $G = 15$~mag, but much poorer at $G = 18.5$~mag. Effective temperature and reddening can be estimated accurately (3--4\% and 0.1--0.2 mag, respectively, at $G = 15$~mag), but there is a strong and ubiquitous degeneracy in these parameters which will ultimately prevent to estimate either accurately at faint magnitudes.

\subsection{Spectroscopy}

In the on-ground data processing, radial velocities will be obtained by cross-correlating spectra with a template \citep{2007sf2a.conf..485G}. An initial estimate of the source atmospheric parameters derived from the astrometric and photometric data will be used to select the most appropriate template. Iterative improvements of this procedure are then made. For stars brighter than $15^{\rm th}$ magnitude, it will be possible to derive radial velocities from spectra obtained during a single field-of-view transit. For fainter stars, down to $17^{\rm th}$ magnitude, only summation of all transit spectra collected during the entire mission will allow the determination of mission-averaged radial velocities. The dependence of the end-of-mission radial-velocity standard error on spectral type and magnitude is shown in Figure~\ref{deBruijne_Figure_RVS_performance} (the error also depends on position on the sky through the number of focal-plane transits over the lifetime; performance numbers reported here refer to a sky average).

For the brightest stars, stellar atmospheric parameters will also be extracted from the RVS spectra, by comparison of the latter to a library of reference-star spectra using minimum-distance methods, principal-component analyses, and neural-network approaches \citep{2011A&A...535A.106K}. The determination of these source parameters will also rely on information collected by the other two instruments: astrometric data will constrain surface gravities, while photometric observations will provide information on many astrophysical parameters, for instance reddening.

\acknowledgments
This paper is partially based on information provided on ESA's Science-and-Techno\-logy and Research-and-Scientific-Support-Department web pages, \url{http://sci.esa.int/gaia} and \url{http://www.rssd.esa.int/gaia}.

\bibliographystyle{spr-mp-nameyear-cnd}
\bibliography{deBruijne}

\end{document}